# Visualizing Three-Dimensional Micromechanical Response in Nanomaterials

**David Bronfenbrenner[1], Matthew Bibee[2], and Apurva Mehta[2]***

[1]*Dept. of Materials Sci and Eng., University of California, Berkeley, CA 94720*
[2]*Stanford Synchrotron Radiation Lightsource, Menlo Park CA 94025*

Keywords: diffraction, nanocrystalline materials, micromechanical deformation, reciprocal space, strain determination

**Abstract**

Understanding mechanical properties of materials requires not only complete determination of the three-dimensional response at a local scale, but also knowledge of the mode or the mechanism by which deformation takes place. Probing mechanical response at such a depth can be only achieved through a diffraction based method. In spite of this, diffraction based methods still are not commonly employed for strain measurements because they are perceived as very time intensive and non-intuitive. Herein we introduce the concept of a diffraction strain ellipsoid, and show how its shape, thickness, and orientation represent the complete deformation state in a powerfully visual and intuitive way. We also show how the geometry of the ellipsoid can be very quickly determined from x-ray diffraction data obtained using a large area detector, and how it can be used to understand micromechanical deformation of nanocrystalline materials.

---

* *Correspondence to:* Apurva Mehta; e-mail: mehta@SLAC.stanford.edu

## *Introduction:*

The primary goal of investigating the mechanical behavior of materials is to understand how they deform under externally applied loads. The resultant deformations are often complex and always three-dimensional, even for a uniaxial applied load. Conventional measurements of these deformation responses rely on physical gauges, which are in contact with the sample and convert change of length into strain. Such methods measure average strain over a large region of the sample and obtain strain states that are partial at best and often only unidirectional. Non-contact optical methods using laser gauges have gone a step further by measuring strain on a smaller scale and in two dimensions but are still unable to probe the complete three-dimensional response. Another "optical" method that is becoming popular is Digital Image Correlation (DIC); which routinely determines two-dimensional deformation at the scale of the optical resolution of the system, and recently there have been a few attempts to use multiple camera systems to probe the deformation in the third dimension [1, 2, 3].

All of these methods, including three-dimensional DIC, however, measure the mechanical response via the displacement of some feature on (or in) the sample. Therefore these techniques measure the gross or "bulk" deformation of the sample (often called macrostrain) and are completely insensitive to the actual mechanism by which the material deforms. This insensitivity to the mechanism by which deformation occurs is a critical failing in understanding mechanical response, as a material can respond to an external load by several very different pathways simultaneously and with dramatically different consequences. For example, the response can arise from reversible distortion or flexure of interatomic bonds. In this case the strain energy is stored in the system, and on withdrawal of the external load the stored energy will tend to reform the material to its original shape. Alternately, the mechanical response can occur through breakage of bonds (e.g. formation of dislocations or slip planes), and the energy driving such a response is dissipated as heat and the response is not reversible. Therefore, one can argue that understanding the pathway by which the material deforms is perhaps even more crucial than knowing the magnitude of the deformation, as a material can withstand large deformations in one mode, but may fail under even a small amount of deformation in another mode.

What is required for a fuller understanding of material response to mechanical impetus is a direct measurement of the local microscopic response (i.e., a true microstrain measurement). Conventional strain measurement, though often a good starting point for such an investigation, is generally insufficient for such a task. The ideal technique should look at the rearrangement of atoms in the material as a response to the external load, which may then be compared to the overall, "bulk" mechanical response. Such a method would allow the researcher to determine both the magnitude and mechanism of deformation. A technique, therefore, that can directly measure the true three-dimensional microstrain response of a material will have big impact on understanding of mechanical behavior of materials.

Determination of strain via diffraction is such a technique (and the only such technique). X-ray, neutron and in some cases electron diffraction have all been used for determination of strain [4, 5, 6, 7, 8] in spite of being a superior method for determination of the mechanical response of a material to applied load, diffraction is not very widely used. We believe there are two operational challenges that limit the application of diffraction methods for strain determination and consequently limit the path to a deeper understanding of material properties and mechanical response.

The first hurdle is experimental; i.e., the traditional implementation of the diffraction method for strain determination is very time intensive. But this barrier has been lowered by the development of fast read-out large area detectors and the proliferation of high brightness synchrotron x-ray sources.

The second hurdle is conceptual; in many formulations the relationship between diffraction space and deformation/strain is buried in mathematics and is not intuitive or visually accessible   In order to make the experiments less time intensive, often several simplifying assumptions are made and limitations imposed. Implications of these restrictions and assumptions on strain determination may not be explicitly stated and give rise to confusion about when a particular formulation of the diffraction method is applicable. For example, use of large area detectors can necessitate changes in the diffraction geometry. This in turn modifies the mathematical formulation and, as will be shown below, the reformulation required is not intuitively obvious. Another common example of the confusion arising from the non-intuitive and visually inaccessible formulation is the lack of understanding of the limitations and assumptions under which $\sin^2\psi$, a restricted formulation of the diffraction method, gives meaningful results.

In this paper, we introduce the concept of the "diffraction strain ellipsoid" and recast the conditions necessary to obtain the three-dimensional elastic response in terms of the data necessary to obtain the shape and orientation of that surface. We will then show how visualization of elastic strain in terms of the diffraction strain ellipsoid can be useful in reformulating the diffraction physics and propose several alternate sets of measurement geometries for obtaining the full three-dimensional elastic response using a large area detector.  We will then show the use of these new geometries (and formulations) to obtain the mechanical response of several nanocrystalline and even amorphous materials. Finally, we will show how the visualization of mechanical deformation in terms of a diffraction strain ellipsoid can be extended even further to represent, and eventually even quantify various other modes of deformation, including dislocation driven plasticity.

From the diffraction strain ellipsoid formulations, it will become evident that one of the better suited geometries for investigation of micromechanical deformation in nanocrystalline inorganic materials requires a multi-wavelength (from 0.5 to 1.5 A) diffractometer capable of producing a small bright spot.  At present it is fairly routine to obtain an incident beam of a few microns in diameter in the $5 - 20$ KeV (wavelength of 0.5 to 1.5 A) at most synchrotron rings; therefore, an instrument on such a beamline dedicated to micromechanical studies will be invaluable. For this reason, we will primarily focus development of an x-ray diffraction technique.  But with some modification, many aspects of the technique developed here can be applied to the task of determination of micromechanical deformation from other types of diffraction experiments, such electron and neutron diffraction, and small angle x-ray and neutron scattering, as well.

## *Methodology:*

### **Probing interatomic spacing via diffraction:**

Diffraction measures the elastic scatter of the incident beam. The change in momentum suffered by the incident radiation must be matched by the change in collective momentum (of the atoms) in the scattering material. The periodicity of atomic arrangement in a crystalline material (or short range order and symmetry of atomic arrangements in poorly crystalline and even amorphous material) results in only a limited set of allowable momentum 'kicks'. Consequently, the incident beam can scatter by only discrete amounts if and only if the scattering crystal is aligned appropriately.

These discrete allowed momentum transfer vectors can be arranged, based on their magnitude and direction, in a three-dimensional construction called a reciprocal lattice, and denoted by the Miller indices hkl for a crystalline material. In this representation of diffraction, an incident beam of radiation is scattered by the crystal if the momentum change suffered by the beam on scatter matches one of the reciprocal-lattice vectors. In elastic scattering there is no change in the magnitude of the momentum, therefore, the change in momentum comes entirely from a change in the direction of propagation. Thus, the end of the momentum vector of all possible elastically scattered (i.e., diffracted) beams defines a spherical surface referred to as the Ewald sphere. The momentum matching condition mentioned above can be graphically expressed by intersection of the Ewald sphere with the reciprocal lattice, as shown in Figure 1a and discussed in many textbooks on diffraction [9, 10].

The reciprocal lattice and the real space interatomic arrangement are related to each other through Fourier transforms. As there is a well defined relationship between these two spaces, when a material distorts under the application of an externally applied load leading to a change in the interatomic arrangement in real space, there is a corresponding well defined distortion of the reciprocal lattice that fully captures every distortion of the interatomic arrangement. Different modes of deformation alter the reciprocal lattice in different ways. For example, dislocation driven plasticity broadens the reciprocal lattice points in specific directions, whereas bond-flexure driven elastic deformation changes the location of the reciprocal lattice points. Measurement of strain via diffraction, therefore, involves measurement of type and magnitude of the reciprocal lattice distortion.

Each crystallite of each phase in a material is represented by a reciprocal lattice. If the material under investigation is either a single crystal or if the crystallite size is bigger than the size of the incident beam (specifically, the volume of the crystallite is bigger than the probed-sample volume) there may be very few (and often no) reciprocal lattice vectors illuminated in a diffraction measurement. Under such conditions, a single wavelength measurement would clearly be insufficient, but if the incident beam has a range of energies (wavelengths), in effect imparting some thickness to the Ewald sphere, then several reciprocal lattice points will be in diffraction condition as shown in Fig 1b. In such a "white/pink" beam Laue diffraction measurement the overall three-dimensional distortion of the reciprocal space, and hence of the three-dimensional deformation of that crystallite, can be deduced from a single diffraction pattern. This technique is the basis of many of the microdiffraction beamlines at synchrotron sources that specialize in measuring elastic lattice strain and is dealt with in detail elsewhere [11].

On the other hand, when the crystallite size is significantly smaller than the beamsize the white beam method described above fails, as under these conditions, the x-ray beam illuminates many crystallites with different orientations simultaneously, and white/pink beam Laue diffraction produces far too many diffraction spots to give rise to an indexable diffraction pattern. (Often when the range of energy is sufficiently large and/or the size of the crystallite sufficiently small the diffraction pattern is a composed of a large number of overlapping patterns and can appear to be monotonically illuminated). Even with the current very bright $3^{rd}$ generation synchrotron x-ray sources it is not possible to obtain an x-ray beam smaller than a 100 nm with sufficient numbers of photons for diffraction. Thus, for complex experiments in micromechanics, a diffraction method is needed to extract the full three-dimensional distortion of a nanocrystalline material quickly   Under these conditions, monochromatic diffraction, i.e., a very thin Ewald sphere, still intersects several reciprocal lattice points, albeit from many different crystallites, and forms the basis of the method for extraction of the full three-dimensional distortion for nanocrystalline materials. The use of monochromatic diffraction for strain determination will be explored in detail below.

**Diffraction strain ellipsoid:**

Both "white/pink" beam and monochromatic diffraction measure micromechanical deformation through detailed probing of the reciprocal space using multiple diffraction events. Whereas "white/pink" beam diffraction creates these diffraction events due to a range of energies in the incident beam, in monochromatic diffraction (often called powder diffraction) they occur because a large number of crystallites, from a subset of crystallites illuminated, satisfy the diffraction conditions as illustrated in the graphical construction shown in Figure 1c. From the many randomly oriented crystallites participating in monochromatic diffraction at a given we can construct an ensemble-reciprocal lattice by superposition. This ensemble reciprocal lattice will look like series of nested spheres as shown in Figure 2a. The intersection of the Ewald sphere with these nested reciprocal lattice spheres gives the condition for "powder" diffraction, commonly represented by Bragg's law. A point x-ray (or a neutron or an electron) detector scans along the surface of the Ewald sphere, and the regions where the reciprocal spheres intersect the Ewald sphere appear as peaks. On the other hand, an area detector samples a section of the Ewald sphere surface in a single exposure and the intersections of the nested-reciprocal spheres appear as segments of conic sections known as Hull/Debye-Scherrer patterns. If the detector face is normal to the incident beam then the conic sections are circles and are often called powder diffraction rings.

Under the application of an external load, as the individual crystallites in the sample distort, the nested hkl-reciprocal spheres formed by the ensemble of the individual crystallites distort as well. If the strain is small and locally continuous then the reciprocal sphere distorts to an ellipsoid as seen in Figure 2b. This ellipsoid has the property that the strain in any direction (for a particular crystallographic lattice planes) is inversely proportional to the length of the vector drawn from the center to the surface in that direction. This ellipsoid, therefore, appears to be related to the reciprocal strain ellipsoid discussed in many of the classical textbook on theory of elasticity [12]. The major distinctions between these two ellipsoids is that the ellipsoid we are considering is distinct for every crystallographic set of lattice planes and is measured in reciprocal lattice units, thus to distinguish it from the other more traditional strain ellipsoids, from henceforth, we will refer to these distorted reciprocal spheres as diffraction strain ellipsoids. This ellipsoidal distortion from sphericity, for a purely elastic deformation, can be represented by the second-rank strain tensor. For an elastically isotropic material, each of the nested reciprocal spheres distorts similarly; however, for elastically anisotropic materials the distortion of the diffraction strain ellipsoid is not only related to the applied load but also to the crystallographic anisotropy. Determination of elastic strain determination via diffraction can then be viewed as determination of the shape and orientation of the diffraction strain ellipsoid. The distortion of the diffraction ring due to elastic strain is often very small and not easily visible, but can be made visually evident by performing a radial-to-polar-coordinate transformation [13] as shown in figures 2c and 2d. Under such a transformation a diffraction ring for an unstrained sample, a straight line, distorts to a sigma shaped wiggle on application of uniaxial load.

The primary computational complexity in determination of strain from diffraction arises from establishing a geometric relationship between the diffraction geometry and the orientation of the strained sample; or in other words, a relationship between the Ewald sphere and the diffraction strain ellipsoid. The Ewald sphere is defined by a coordinate system aligned to the incident beam and the principal axes of the diffractometer. The diffraction strain ellipsoids (or the reciprocal spheres) are defined on a coordinate system with the origin at the location of the intersection of the unscattered x-ray beam with the Ewald sphere and the axes aligned with the principal axes of the sample. Often one axis is normal to the sample surface and the other two are orthogonal and lie in the plane of the sample surface. One of these axes is often along the principal strain direction and the other two orthogonal to it., but in any case they are seldom aligned with the axes of the diffractometer reference frame and the Ewald sphere.

The form of the mathematical relationship used to orient the diffraction strain ellipsoid to the Ewald sphere depends on the manner in which the data is collected. Traditionally when data is collected using a point detector, the ellipsoid is defined in terms of the longitude, $\phi$ (the rotation of the diffraction vector about the sample normal), and latitude, $\psi$ (the angle of the diffraction vector relative to the sample normal). As is evident, for a complete determination of the shape of the elastically strained ellipsoid, one requires at least two independent slices, or sufficiently large arcs through the diffraction strain ellipsoid. (Often, these are $\psi$ slices taken at constant $\phi$ orientation.) Point by point determination of the shape of the diffraction strain ellipsoid is very time consuming and very frequently the measurements are limited to just one single $\psi$ slice (often at $\phi = 0$) [4]. By limiting the measurement to a single slice through the ellipsoid, one can obtain no more than 3 of the 6 strain components without further assumptions and simplification. The common assumptions made to obtain the rest of the components are radial symmetry of the ellipsoid around the strain axis –i.e., isotropic Poisson's ratio - and absence of shear components. This truncated measurement approach is the traditional $\sin^2\psi$ technique, a technique that is at the heart of many diffraction based strain measurements, but which fails when the sample is subjected to multi-axial loading, thus limited in application to investigation of simplified model systems.

If, on the other hand, the diffraction data is collected using a large area detector it is possible to obtain a complete slice through the diffraction strain ellipsoid in a single exposure (which often takes seconds and usually takes no more than a few minutes).

Unfortunately, however, the slice obtained in a single exposure is not either a $\psi$ or even a $\phi$ slice, and therefore, for easier ellipsoid reconstruction from area detector "slices" it is convenient to recast the geometric relationship between the Ewald sphere and the diffraction strain ellipsoid in terms of a second set of angles, namely $\chi$, $\theta$, and $\alpha$. Here $\theta$ is the Bragg angle, $\chi$ is the angle between the diffraction plane and one of the principal axes of the diffractometer (often the vertical), and $\alpha$ is the angle between the projection of one of the principal sample axes into the diffraction plane and the incident beam. (Though it is convenient to define the principal sample axes as the principal axes of the diffraction strain ellipsoid, it is not necessary. Any other choice of the sample axes can always be related to the principal axes of the diffraction strain ellipsoid through invariants of strain [12]). The relationship between the angles used in the point detector and area detector configuration is illustrated in Figure 2e.

The relationships between the Ewald sphere – diffraction space – and the two different orientation bases for the diffraction strain ellipsoid in terms of the components of the 2$^{nd}$ rank elastic strain tensor can be written in general as follows [14],

$$\varepsilon_Q = \varepsilon_{xx} r_x r_x + \varepsilon_{yy} r_y r_y + \varepsilon_{zz} r_z r_z + \varepsilon_{xy} r_x r_y + \varepsilon_{yz} r_y r_z + \varepsilon_{xz} r_x r_z \quad (1.1)$$

Where $\varepsilon_Q$ is the strain along the diffraction vector and $r_i$ are the rotations associated with the defining diffraction geometry. Using the latitude and longitude geometry (i.e., geometry commonly used with a point detector), this equation becomes

$$\begin{aligned}\varepsilon_{\phi\psi} =\ & \varepsilon_{xx} \cos^2\phi \sin^2\psi + \varepsilon_{yy} \sin^2\phi \sin^2\psi + \varepsilon_{zz} \cos^2\psi + \\ & \varepsilon_{xy} \sin 2\phi \sin^2\psi + \varepsilon_{yz} \sin\phi \sin 2\psi + \varepsilon_{xz} \cos\phi \sin 2\psi\end{aligned} \quad (1.2)$$

This is the common equation for the generalized "sin$^2\psi$ technique" and is often shown in texts such as Noyen and Hauk [4, 5].

When using an area detector, the equation looks slightly more complicated but the angles themselves are easily measured from a powder diffraction ring pattern.

$$\begin{aligned}
\varepsilon_{\theta\chi\alpha} = \ & \varepsilon_{xx}\left(\sin\theta\cos\alpha - \cos\chi\cos\theta\sin\alpha\right)^2 + \\
& \varepsilon_{yy}\sin^2\chi\cos^2\theta + \\
& \varepsilon_{zz}\left(\sin\theta\sin\alpha + \cos\chi\cos\theta\cos\alpha\right)^2 + \\
& \varepsilon_{xy}\left(\sin\chi\sin 2\theta\cos\alpha - \sin 2\chi\cos^2\theta\sin\alpha\right) + \\
& \varepsilon_{yz}\left(\sin\chi\sin 2\theta\sin\alpha + \sin 2\chi\cos^2\theta\cos\alpha\right) + \\
& \varepsilon_{xz}\left(\sin 2\alpha\left(\sin^2\theta - \cos^2\chi\cos^2\theta\right) + \cos\chi\sin 2\theta\cos 2\alpha\right)
\end{aligned} \quad (1.3)$$

## Measuring the Diffraction Strain Ellipsoid Using an Area Detector:

A powder ring on a two-dimensional detector represents a slice of the strain ellipsoid, revealing the variation of lattice spacing over a full range of the angle $\chi$. There are six unknowns in equation 1.3 (or for that matter equation 1.2), whereas each powder ring provides only five quantities that vary significantly (i.e., cos $\chi$, sin $\chi$, cos$^2$ $\chi$, sin$^2$ $\chi$, and sin $2\chi$). Therefore, one diffraction ring from a single diffraction pattern is not sufficient to uniquely determine the full three-dimensional elastic response. In a graphical representation the above discovery translates again, as discussed above, into the need for two independent slices of the diffraction strain ellipsoid to determine its shape in three dimensions.

There are two approaches to address the insufficiency of data obtained from just one 2D diffraction ring to solve Eq 1.3. The first approach is to make a simplifying assumption about the load geometry and the material properties of the sample and eliminate one or more of the independent components of the 2$^{nd}$ rank strain tensor from Eq 1.3. This is the approach proposed by He and Smith [15] and Almer and Stock [16]. The most common simplification is to assume that the sample is elastically isotropic, and the applied load is uniaxial. Under this assumption, the three-dimensional response reduces to two dimensions, and Eq 1.3 can be solved from a single diffraction pattern. (In effect, this assumption states that the Poisson ratio is isotropic.) Many materials of technological importance, such as Cu, and brass are, however, highly anisotropic [17]. For highly anisotropic materials Poisson's ratio can still be isotropic if the sample exhibits no significant crystallographic texture at hierarchically finer local scale, and the assumption of transverse isotropicity can be still used to reduce the problem of 3D strain determination to 2D. But unfortunately, many of the processes for producing technological components involve rolling, drawing or some deposition process that results in highly textured materials. This is also true of many biomaterials, such as nacre, horn, and bone. Therefore, for many of the interesting and technologically important nanocrystalline material and components subjected to complex multi-axial loads there is no alternative but to obtain at least two independent slices of the diffraction strain ellipsoid for determination of the full 3D strain response. From equation 1.3 it is evident that there are only two ways of obtaining a second independent slice of the diffraction strain ellipsoid; either by changing the sample orientation ($\alpha$) or by altering the Bragg angle ($\theta$). Below we will explore three different data collection strategy that vary either $\alpha$ or $\theta$ to obtain at least two independent slices of the diffraction strain ellipsoid using a large area detector.

The first of these strategies is to tilt the sample relative to the incident beam about a single axis, preferentially perpendicular to the incident beam, thus changing α to obtain the second slice through the ellipsoid. For statistical reasons, several different tilt angles are used and the diffraction strain ellipsoid is traced out as seen in Figure 3a. This technique is universal as it is viable for any type of diffraction, wide or small angle, and for incident radiation from any energy source (i.e. x-rays, electron, neutrons) since the sample orientation to the beam does not affect the diffraction characteristics. The problems with the sample tilting method, however, arise from the change in spot size on the sample surface. Varying beam footprint causes problems if the strain inhomogeneity is comparable to the spot size., Further problems are caused if the sample is not very precisely aligned with the rotation axis, as then when the sample is rotated not only does the beam spread out on the sample but it also moves to a different location. Therefore, though this technique is universal, it is liable to cause difficulties with samples that have complex strain distributions.

For samples with complex strain distributions, where change of the beam footprint associated with varying the sample tilt must be avoided, multiple slices through the diffraction strain ellipsoid still can be obtained, provided that the size of the Ewald sphere can be altered significantly in comparison with the size of the diffraction strain ellipsoid (i.e., Bragg angle, θ, is not too small). Under these conditions the tilt angle, α, and thus the size and location of the beam footprint, can be kept constant and the Bragg angle, θ, can be changed by varying the energy (wavelength) of the incident beam. As the photon energy is varied the size of the Ewald sphere changes and the intersection with the diffraction strain ellipsoid provides an additional independent slice. In this way, several different slices of the diffraction strain ellipsoid can be obtained as shown in Figure 3b. This method works very well for any material that has a well defined and strong diffraction ring between 1-6 Å measured in a diffractometer with reasonably curved Ewald's sphere (i.e., with a wavelength of ~ 0.5 to 1.5 A). This procedure will also work with thermal neutrons, if the wavelength of the neutron beam can be altered. However, it fails if the incident beam wavelength is too small and the Ewald sphere is relatively flat so change in energy does not significantly change the Bragg angle, $\theta$, to yield a statistically different slice of the diffraction strain ellipsoid as would be the case with high energy electron or x-ray diffraction.

Both of the aforementioned methods require at least two diffraction patterns to map out the diffraction strain ellipsoid. But under certain conditions, it is possible to use multiple rings from a single diffraction pattern to determine the full strain tensor. If the wavelength of the incident beam is sufficiently small, but not so small as to make Bragg angles very small, then it is possible to obtain sets of crystallographically related rings (e.g., 100 and 200) at sufficiently different Bragg angles to enable mapping of the diffraction strain ellipsoid in a manner reminiscent of the variable energy method. This strategy of using higher order diffraction rings to vary the Bragg angle and thus trace out the shape of the diffraction strain ellipsoid can be also applied to crystallographically dissimilar rings (i.e., 100 and 110) if the material is known to be isotropic or the crystallographic elastic anisotropy for those rings are previously determined for a material with similar grain size and texture - perhaps by application of one of the multi-diffraction pattern based methods. (For example, if the different diffraction strain ellipsoids in Figure 2b are from crystallographically related rings, they act as three different Bragg angles for a single diffraction strain ellipsoid as seen in Figure 3b. This idea is illustrated in Figure 3c.)

At least one of the strategies described above can be used to obtain strain from any diffraction measurement, be it electron, neutrons or x-ray diffraction. But we believe that a diffractometer at a synchrotron x-ray source with an energy range of 6-40 keV and a small spot size, a robust load frame, and a large area detector, is the ideal instrument for studies of micromechanics in nanocrystalline materials, as on such an instrument all three of the strategies described above can be easily attempted.

## *Experimental Details:*

For testing some the strategies proposed above, we used three nanocrystalline metals, BCC-Fe, FCC-Fe (Stainless Steel) and HCP-Ti thin films with thicknesses of ~25 μm, and a thin polymer film of kapton of ~200 μm thickness. The metallic dogbone samples were laser machined and each sample was painted with a $TiO_2$ paint that provided an internal calibration (and zero strain reference) for every diffraction pattern obtained. Dogbone samples from Kapton were painted with $LaB_6$ powder for internal calibration. The samples were loaded in a custom designed *in-situ* strain rig. One of the grips of the rig was attached to a force transducer and the other grip was driven by a computer control precision motor (Newport LTA-HL motorized actuator). The force and displacement were measured through-out the experiment and a "bulk" or "rig" stress-strain relationship was obtained from those measurements. For diffraction measurements, the samples were loaded close to the yield point and then unloaded at a displacement rate of 50 nm/s until a desired load drop was achieved, and then the diffraction data was collected at that displacement. After data collection, the sample was unloaded again to a lower load and the sequence was repeated until the sample was fully unloaded. Diffraction patterns were collected on beamline 11-3 at the Stanford Synchrotron Radiation Lightsource, at a fixed x-ray energy of 12.7 keV ($\lambda = 0.976$Å). Data was collected on the unloading curve to not only minimize the slippage at the grips, but also to ensure that the deformation of the sample was as completely elastic as possible.

## *Results and Discussion:*

### **Elastic Deformation:**

Thin film samples made from BCC-Fe, HCP-Ti and Kapton were used to verify the "tilting technique".

For BCC-Fe, the full $2^{nd}$ rank strain tensor was determined along the 110, 200 and 211 crystallographic directions at each unloading point. Figure 4 shows one of the cuts (at $\alpha = 0$) through the diffraction ellipsoid and the fit though it for the three crystallographic directions. Average stress at each point was calculated based on the applied load and the measured average cross-section of the dog-bone. The stress and three of the six components of the $2^{nd}$ rank elastic tensor are plotted in Figure 5. The other three components were also obtained, but sake of clarity they were omitted from the Figure. (The statistical errors associated with the all strain calculations for each diffraction point are on the order of +/- 0.001% strain and are too small to be shown in Figures 5 and 6.) The measured shear components obtained are not significantly different from zero, indicating that the applied external load was predominatly uniaxial along one of the principal sample axes. We also calculated the average strain (macrostrain) from the crosshead displacement of the moving jaw of the rig, and plot the macro-stress-strain relation for the unloading curve as the bold ochre solid line. The Poisson ratio for the three crystallographic directions as a function of load is shown in Figure 5 (inset) Our measurements show that the Poisson ration becomes approximately 1/3 ( ~0.33 +/- 0.02) at the larger stress values for all three crystallographic directions, as would be expected for most metallic materials [17].

Further, our diffraction measurements show that the crystallographic elastic anisotropy can be measured even in a polycrystalline sample: an impossible task by any other strain measurement technique. Our measurements on nanocrystalline bcc Fe indicates that the stiffer directions (110, and 211) of the materials have elastic moduli insignificantly different from those in single crystal (Reuss value of the moduli is 210 GPa for each), but the softer (200) crystallographic direction has becoming significantly stiffer (167 GPa measured as compared to Reuss value of 125 GPa). We observed similar stiffening of the softer direction in other anisotropic nanocrystalline metals, such as Ag and Cu. Therefore, our preliminary measurements of the elastic anisotropy in polycrystalline metals suggest that the apportioning of the applied load (and thus local stress) is highly anisotropic along different crystallographic directions. The siffer crystallographic directions bear load nearly equivalent to the average applied load, whereas the softer crystallographic directions bear significantly less. These results then are an excellent example of the insight into micromechanical deformation that can only be obtained from determination of the microstrain response, made possible by diffraction measurements.

As mentioned earlier the "tilting" technique using a large area detector can be applied to any material exhibiting diffraction, including polymeric materials showing intermediate-range order such as Kapton. Figure 6a demonstrates this by comparing the strain along the loading direction obtained from the "tilting" techniques applied to one of the diffuse diffraction rings from the Kapton diffraction patterns ($\alpha$ of -40°, -20°, 0°) to the strain obtained from the crosshead displacement unloading curve. (The technique yields the full 2$^{nd}$ rank strain tensor, as shown above. But for the sake of clarity we have not shown the other 5 components.)

The HCP-Ti was used to compare the strain tensors obtained by two different strategies described above. We applied the "tilting" technique to obtain the strain in the 101 and 202 and several other crystallographic directions. (We used three tilts; $\alpha$ of -20°, 0°, 15°) We then chose a single tilt angles and applied the strategy of using multiple Bragg angles ($\theta$) obtained from crystallographically related higher order diffraction rings (i.e., set of 101 and 202 rings), to solve for the shape of the diffraction strain ellipsoid. The single shot method is experimentally most attractive as it requires a single diffraction pattern to obtain the full 2$^{nd}$ rank elastic strain tensor, though its scope is limited by the ability to obtain crystallographically related higher order rings with sufficiently different Bragg angles (and without overlaps with other unrelated rings). In Figure 6b, we compare strain along the loading direction thus obtained from a single diffraction pattern ($\alpha = -20°$) to the strain obtained from the "tilting technique" using three diffraction patterns at three different tilt angles. The data shows that the strain is equivalent whether it is calculated using the tilting method or the single shot method using higher order diffraction rings.

**Other Modes of Deformation:**

So far, in this paper, we have implicitly treated deformation of a material in an almost purely elastic regime, i.e., from flexure of bonds only. Under these conditions the reciprocal sphere deforms into a perfect ellipsoid and can be adequately described by six parameters, namely the lengths of the three principle axes and its orientation in space (i.e., its tilt, pitch and roll with respect the laboratory frame of reference). These six parameters can be transformed into the six conventional strain parameters via equation 1.3, as shown above.

However, materials do not only deform via elastic flexure of bonds, but routinely respond to the applied load through many other mechanisms, including a first order thermomechanical phase transition or breakage of bonds creating line and plane defects along which the lattice deforms and slides. Often these other modes of deformation are of higher interest in a micromechanical investigation as they drive unusual mechanical properties, such as superelasticity, or are responsible for irreversible distortions and ultimately failure.

Because a reciprocal lattice is a Fourier transform of the actual arrangement of atoms in a material, it must therefore contain information about every possible distortion of the perfect lattice induced by all possible modes of deformation a material can undergo. Fortunately, all of these different modes of deformation distort the reciprocal lattice in a distinctly different manner. Therefore, diffraction measurements are not only sensitive to these other modes of deformation – they are uniquely placed to capture them, especially if more than one is active simultaneously, as is often the case

In Figure 7a, we show deformation of nanocrystalline stainless steel in the nominally elastic regime (i.e., in unloading just before significant irreversible deformation has occurred). The diffraction signature of deformation is very different from that for BCC-Fe as shown in Figure 2c and d. In fact, the deformation surface is not a true ellipsoid at all, but perhaps can be visualized as an ellipsoid with pop-backs and gouges. Figure 7b shows the data plotted in radial coordinates while Figure 7c is a schematic recreation of a possible reciprocal strain surface based on partial sampling of the surface via the tilting technique. We have noticed similar distortions in other "soft" materials such as Al and Cu as well. We believe these distortions are an indication that these soft materials, even when subjected to loading in the nominally elastic regime, are deforming by multiple modes. We believe that alternative modes are so easily activated because these face-centered materials have access to several very easy slip systems. Deformation of a material via slippage along certain crystallographic planes relieves the built-up elastic energy along those directions, and the elastically deformed reciprocal sphere (i.e., the diffraction strain ellipsoid) experiences "pop-backs" towards the undeformed reciprocal sphere along those directions. Deformation through slippage along specific crystallographic planes and directions is then seen as a gouge or a ridge on the diffraction strain ellipsoid, depending on the whether the strain is compressive or tensile at the location of the pop-back.

Other effects are also observed – for instance, dislocation driven plasticity tends to break up the long range coherence of the crystal lattice, and as a result appears as broadening of the diffraction spots (or the diffraction ring, in the case of powder diffraction) along certain crystallographic directions. This can be seen as a thickening of the diffraction strain ellipsoid in the orientations undergoing dislocation driven plasticity. Another common mode of plasticity is deformation that rotates and reorients certain grains, which can be also thought of as a change in crystallographic texture of the sample. Such a reorientation will redistribute the density of the diffraction strain ellipsoid. We have observed that many materials in predominantly plastic deformation regime often exhibit all of these modes simultaneously.

Another not so common but very interesting mode of strain accommodation in a crystal is through a first order thermomechanical phase transition that is at the heart of superelasticity and shapememory in materials like NiTi. When a material undergoes a stress driven phase transition, the diffraction strain ellipsoid of the parent phase loses weight in certain easy transformation directions and reforms into two or more new diffraction strain ellipsoids corresponding to the transformed phase and possessing different eccentricity, weight and average diameter. In our micromechanical deformation of nanocrystalline NiTi we have seen all of these four (two reversible elastic and two irreversible) modes of lattice deformation in diffraction patterns.

All of these other modes of deformation distort a reciprocal sphere in very specific ways, each more complex than an ellipsoidal deformation arising from purely reversible bond flexure, and clearly can not be represented by just 6 parameters used to define the shape of a perfect ellipsoid (or for that matter the $2^{nd}$ rank strain tensor). Often, as seen in the case of stainless steel, material deforms via several of these modes simultaneously. Both of these observations indicate that for a fuller understanding of micromechanical deformation we need a more complex mathematical formulation than that used to define the surface and orientation of the diffraction strain ellipsoid (and the $2^{nd}$ rank strain tensor currently used to describe it). Nevertheless, visualization of the deformation of a material in terms of shape, thickness and distortion of diffraction strain ellipsoids, we believe, is a very convenient, concise method for capturing the full micromechanical state of deformation, at least qualitatively. Thus, the concept of a diffraction strain ellipsoid that can fully encapsulate the density, thickness, local (i.e., "pop-back" and gouges) and global (i.e., ellipsoidal eccentricity) nature of the reciprocal strain surface can still serve a very useful purpose as a foundation for a complete mathematical description of the micromechanical strain state of a polycrystalline material.

## *Conclusions:*

In conclusion, we show how diffraction and especially x-ray diffraction using a large area detector can not only measure the full three-dimensional mechanical response of a material, but also can distinguish among various microscopic modes by which deformation occurs. Diffraction is, therefore, an ideal tool for studies of micromechanics.

We also introduce the concept of the diffraction strain ellipsoid and show how the physics of diffraction involved in micromechanical strain determination can be made visual and transparent by transposing the problem of strain determination into a determination of the shape, orientation, local distortion and thickness of the diffraction strain ellipsoid. Such a visualization of the micromechanics of deformations leads to not only a robust determination of the elastic deformation, but provides an easy and intuitive description and perhaps even a method of quantification for the other, still poorly characterized, modes of deformation.

## *Acknowledgments:*

We would like to thank Alan Pelton and Monica Barney for help with sample preparation and Synchrotron data collection. Diffraction data shown here was collected at Stanford Synchrotron Radiation Lightsource, which is supported by the U.S. Department of Energy, Office of Science, Office of Basic Energy Sciences,

## References:

[1] Bruck HA, McNeill SR, Sutton MA, and Peters WH, *Exp. Mech.* 1989;29(3):261.

[2] Cheng P, Sutton MA, Schreier HW, and McNeill SR, *Exp. Mech.* 2002;42(3):344.

[3] Verhulp. Van Rietbergen EB, and Huiskes R, *J. Biomech.* 2004;37(9):1313.

[4] Noyen IC and Cohen JB. Residual Stress Measurement by Diffraction and Interpretation. Springer-Verlag, New York, 1987.

[5] Hauk V. Structural and Residual Stress Analysis by Nondestructive Methods. Elsevier, Amsterdam, 1997.

[6] Hutchings MT, Withers PJ, Holden TM, Lorentzen T. Introduction to the Characterization of Residual Stress by Neutron Diffraction. Taylor & Francis, Boca Raton, 2005.

[7] Reimers W, Pyzalla AR, Schreyer A, Clemens H. Neutrons and Synchrotron adiation in Engineering Materials Science. Wiley-VHC, Weinheim, 2008.

[8] Fultz B, Howe JM. Transmission Electron Microscopy and Diffractometry of Materials. Springer-Verlag, Berlin, 2008.

[9] Warren BE. X-Ray Diffraction. Dover Publications, Dover, 1990.

[10] Culity BD. Elements of X-ray Diffraction (2 ed.). Addison-Wesley, Reading, MA, 1978.

[11] Tamura N, MacDowell AA, Spolenak R, Valek BC, Bravman JC, Brown WL, Celestre RS, Padmore HA, Batterman BW and Patel JR, *J. of Synchrotron Radiation* 2003;10:137.

[12] In reality the transformation is more complex as the area detector is never perfectly normal to the incident beam, but several software packages, such as the Area Diffraction Machine (http://code.google.com/p/areadiffractionmachine/) or Fit2D (http://www.esrf.eu/computing/scientific/FIT2D/) can easily handle it.

[13] Love, AEH, *Treatise on the Mathematical Theory of Elasticity*. (4 ed.). New York: Dover Publications, 1993.

[14] Bronfenbrenner D. "The Development of an X-Ray-Diffraction Strain-Measurement Technique and its Application to Micromechanical-Deformation Modes." Ph.D. Thesis, U. Calif. Berkeley, 2008.

[15]    He BB and Smith KL. Fundamental equation of strain and stress measurement using 2D detectors. In: Proceedings of the Society of Experimental Mechanics_ Spring Conference on Experimental Mechanics, Society of Experimental Mechanics, Houston, 1998 p. 217–220.

[16]    Almer J and Stock SR, *J. Struct. Biol.* 2005;152:14.

[17]    Hertzberg RW. Deformation and Fracture Mechanics of Engineering Materials. John Wiley and Sons, New York, 1989.

## *Figure Captions:*

Figure 1: (a) The Ewald Sphere construction for a monochromatic incident beam on a single crystallite. (b) When the incident beam is polychromatic (white), the Ewald sphere has multiple radii and many reflections from the single crystallite are seen. (c) When the sample is polycrystalline, the reciprocal lattice occurs in all rotations about a given point and it resembles a circle (or sphere in three dimensions) and multiple reflections can be seen with a monochromatic incident beam.

Figure 2: (a) The intersection of the Ewald Sphere with the nesting reciprocal spheres is shown as rings. Each reciprocal sphere is defined by a different crystallographic direction. (b) Under load, the reciprocal spheres distort into diffraction strain ellipsoids (exaggerated). Note the changes in the diffraction rings. (c) A polar plot of an unstrained diffraction ring. (d) A polar plot of a strained diffraction ring. (e) The sample axes ($S_i$) are in black, the incident beam axes ($I_i$) are in gray, and the diffraction vector is a red line with a red sphere. Each point on a diffraction ring represents a diffraction vector which is defined by the angles $\phi$ and $\psi$, or the angles $\chi$, $\theta$, and $\alpha$.

Figure 3: (a) If multiple tilts are performed, the diffraction strain ellipsoid is traced out in slices as shown. (b) Varying the energy results in slices in this fashion. (c) If the elastic anisotropy is known, rings from several crystallographic orientations may be "collapsed" onto a single diffraction strain ellipsoid.

Figure 4: Polar plots of three rings from uniaxially strained BCC-Fe, at varying load values. Red dots indicate diffraction data. Blue lines show fit of data to a diffraction strain ellipsoid using equation 1.3.

Figure 5: The stress-strain-elastic-unloading curve for BCC-Fe along three different crystallographic directions compared to the bulk behavior. (inset) The calculated Poisson's ratio for the three crystallographic directions using the in-plane orthogonal axis and the unloading axis.

Figure 6: (a) The use of the tilting technique on Kapton showing the effectiveness determining long-range-order strain. (b) The comparison of the tilting method to the single shot method in HCP-Ti.

Figure 7: (a) A stainless steel diffraction ring showing non-elastic deformation behaviour. (b) The same ring plotted in radial coordinates and exaggerated to show the distortions. (c) A possible three dimensional strain surface from a material undergoing non-elastic deformation behaviour. Notice the deviation from a perfect ellipsoid (which would occur for perfectly elastic behaviour.)

*Figures:*

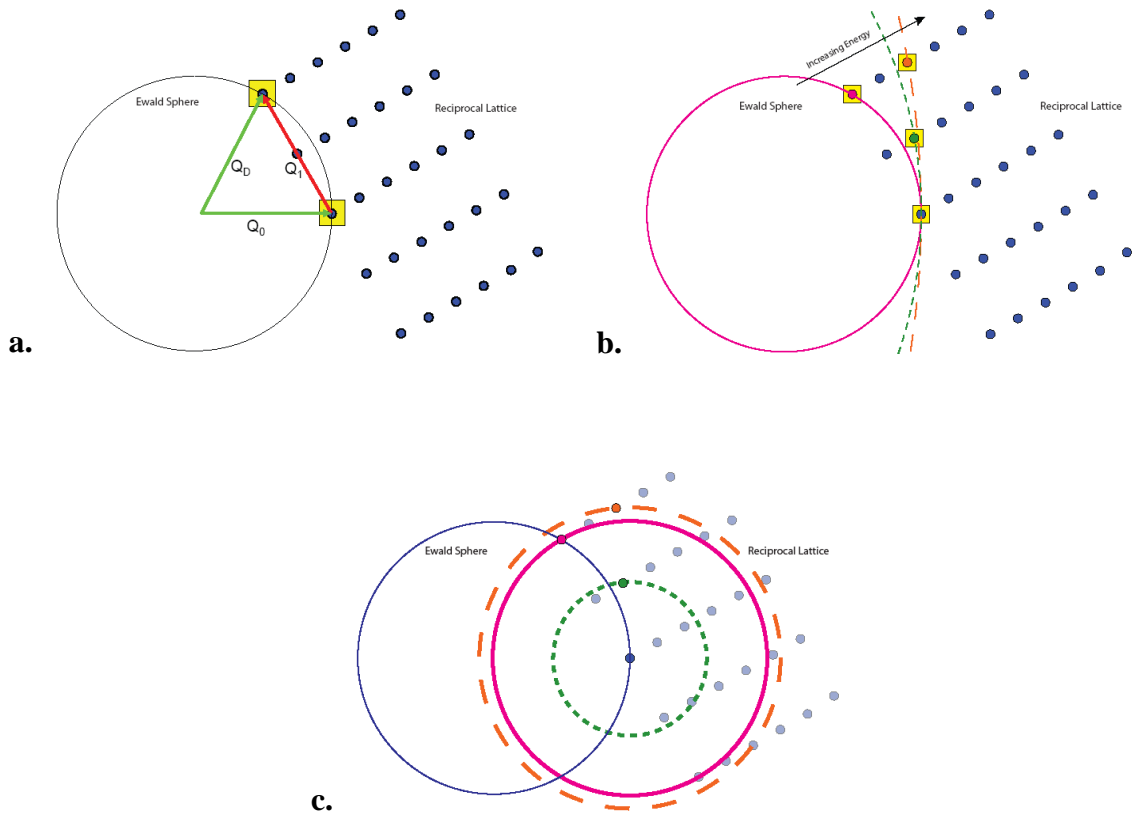

**a.** **b.**

**c.**

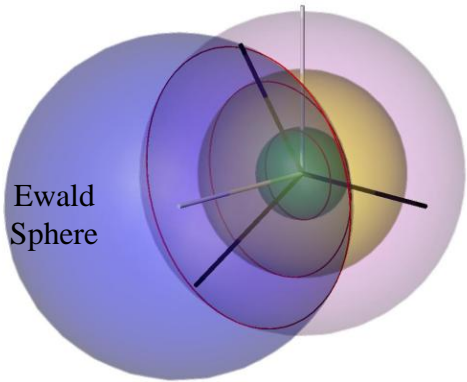 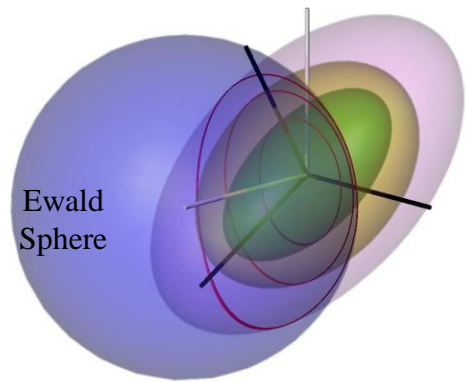

a. b.

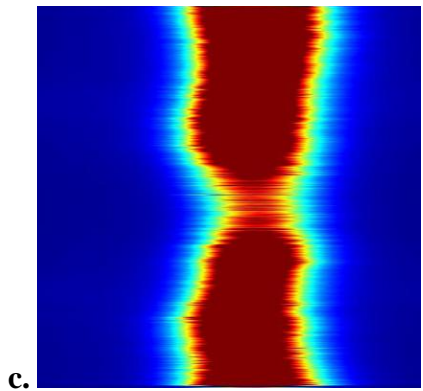 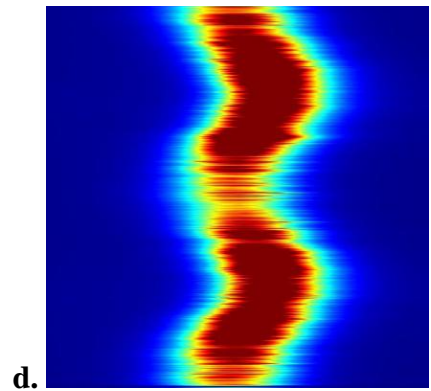

c. d.

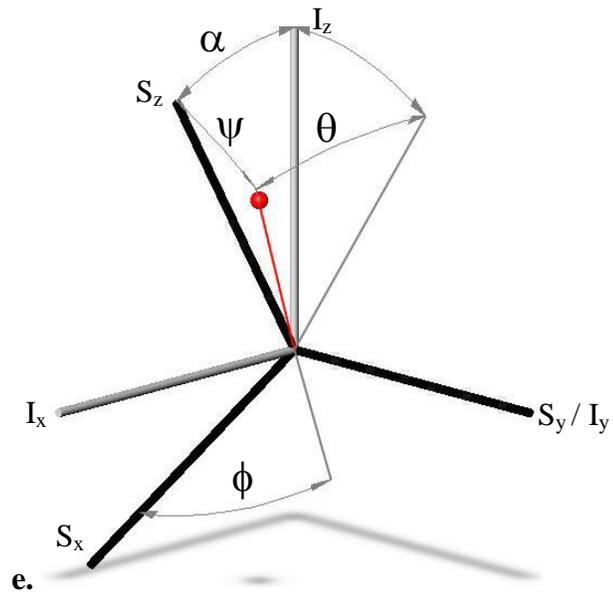

e.

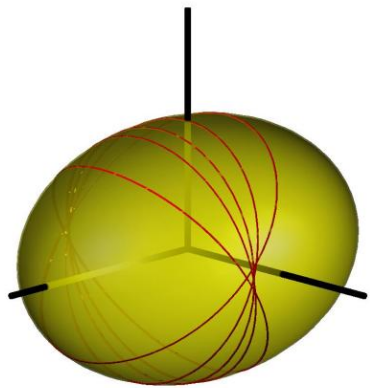

**a.**

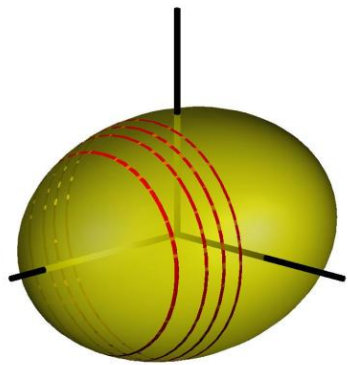

**b.**

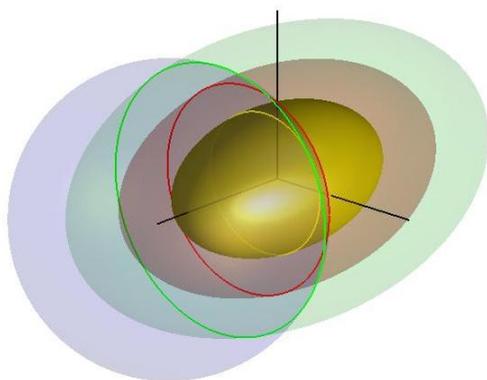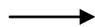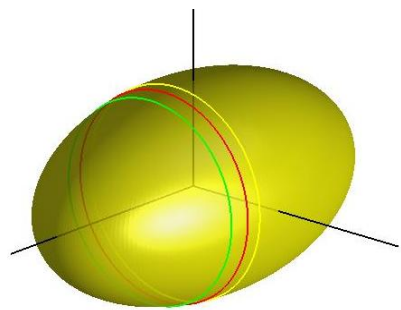

**c.**

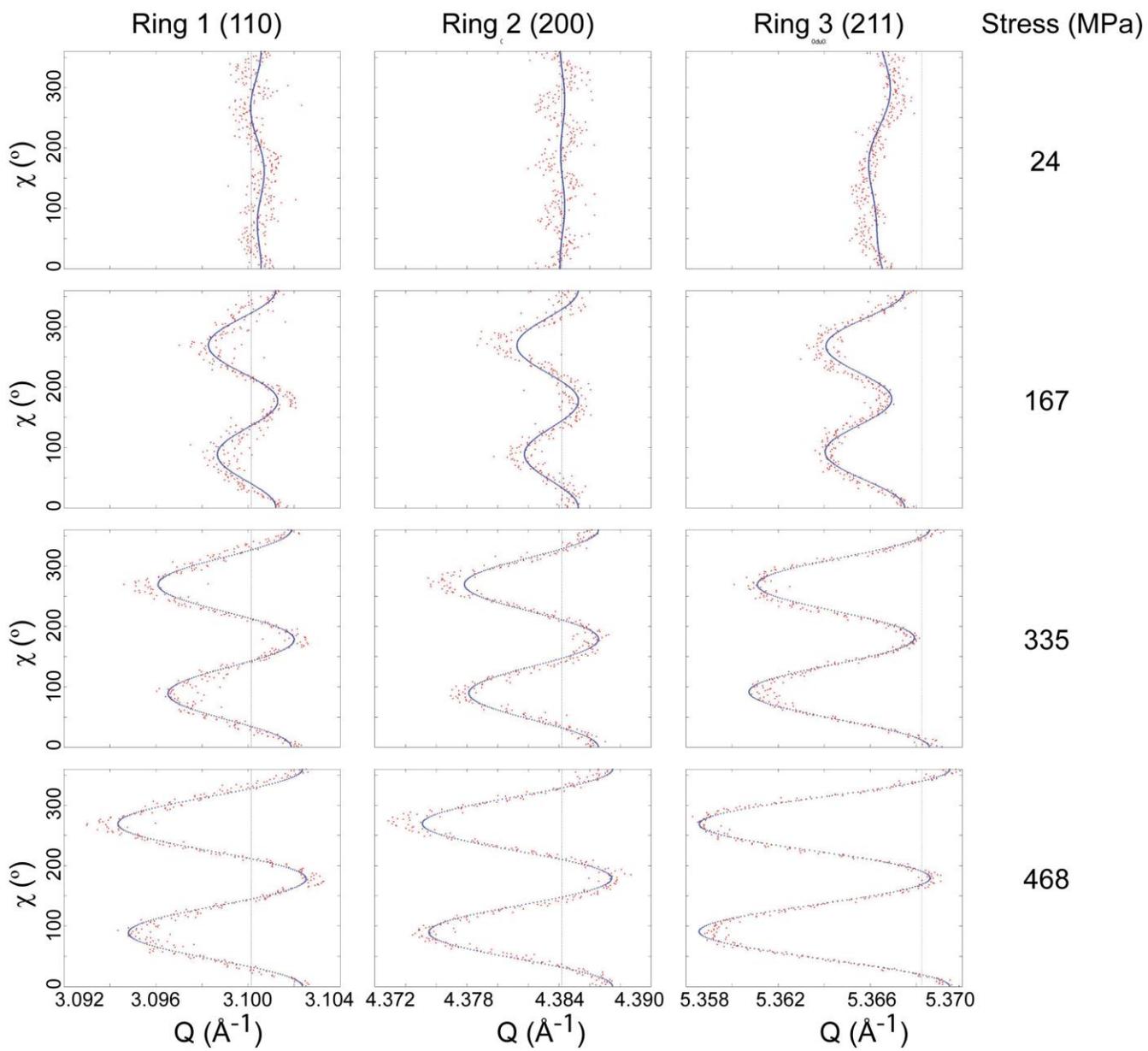

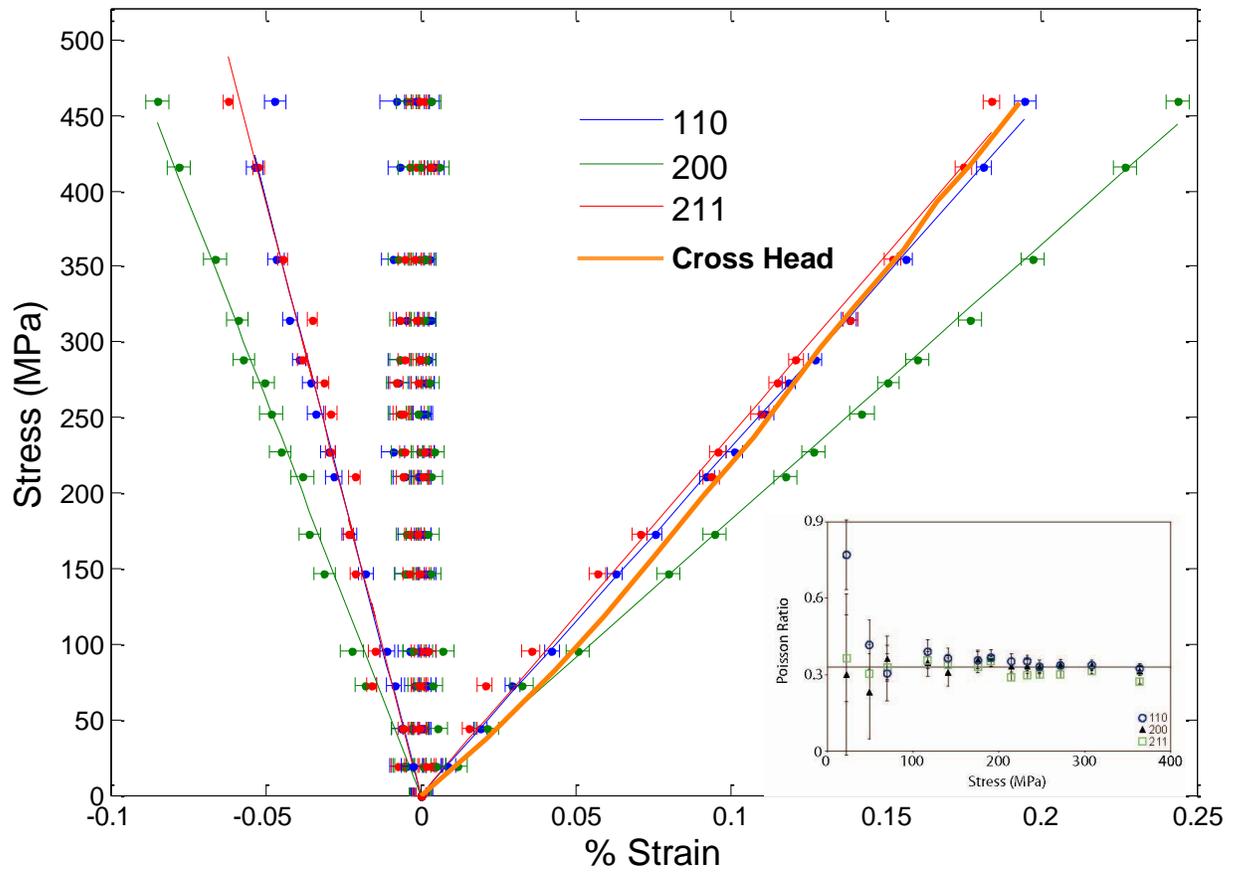

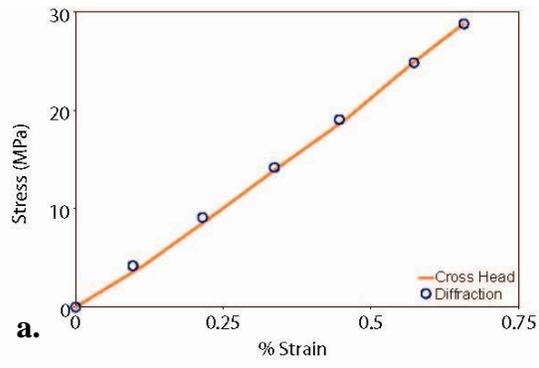 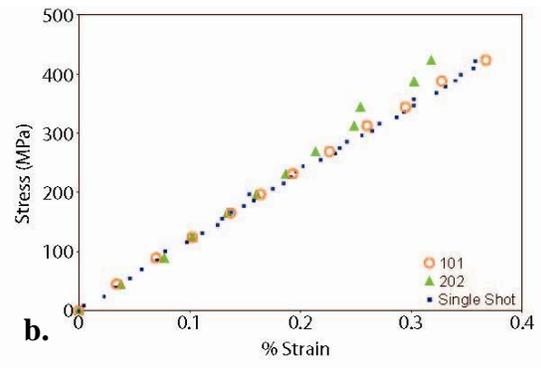

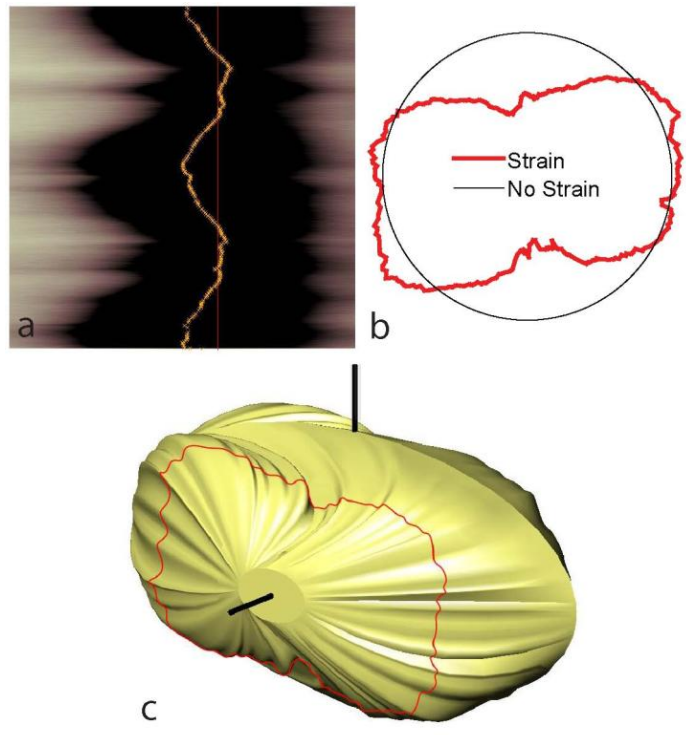